\documentclass{achemso}
\SectionNumbersOn
\usepackage{amsmath}
\usepackage{amssymb}
\usepackage{color}
\usepackage{graphicx}
\usepackage{subcaption}
\graphicspath{{image/}}
\allowdisplaybreaks
\usepackage{ulem}
\usepackage[title]{appendix}

\makeatletter

\renewcommand{\Roman}[1]{\expandafter\@slowromancap\romannumeral #1@}
\makeatother

\title{Gaussian and plane-wave mixed density fitting for periodic systems}
\author{Qiming Sun}
\affiliation{Division of Chemistry and Chemical Engineering, California Institute of Technology, CA 91125}
\email{osirpt.sun@gmail.com}
\author{Timothy C. Berkelbach}
\affiliation{Department of Chemistry and James Franck Institute, University of Chicago, Chicago, IL 60637}
\author{James D. McClain}
\affiliation{Department of Chemistry, Princeton University, Princeton, NJ 08543}
\author{Garnet Kin-Lic Chan}
\affiliation{Division of Chemistry and Chemical Engineering, California Institute of Technology, CA 91125}
\email{gkc1000@gmail.com}

\begin{document}
\maketitle
\begin{abstract}
  We introduce a mixed density fitting scheme that uses both a Gaussian and a plane-wave fitting basis to accurately evaluate electron repulsion integrals in crystalline systems. We use this scheme to enable efficient all-electron  Gaussian based periodic
  density functional and Hartree-Fock calculations.
\end{abstract}

\section{Introduction}
Computing the two-electron repulsion integrals (ERIs)
\begin{equation}
  (\mu\nu|\kappa\lambda)
  = \int \mu^*(\mathbf{r}_1) \nu(\mathbf{r}_1)\frac{1}{r_{12}}
  \kappa^*(\mathbf{r}_2)\lambda(\mathbf{r}_2) d\mathbf{r}_1 \mathbf{r}_2
  \label{eq:eri}
\end{equation}
has been a traditional  bottleneck in electronic structure modeling when using 
 a Gaussian basis. 
The ERIs serve both as final targets of computation,
or may be used in contractions to form intermediates, such as in the Coulomb and exchange operators
in mean-field calculations.

Various approximations have been proposed to reduce the cost of ERI computation and their
associated intermediates in both molecular and crystalline systems.
Many of them, including 
Gaussian density fitting\cite{Whitten1973,Dunlap1979,Vahtras1993,Eichkorn1995,Dunlap2000},
Cholesky decomposition\cite{Beebe1977,Roeeggen1986,Aquilante2009}, 
plane-wave Fourier transform techniques\cite{Fuesti-Molnar2002,Fuesti-Molnar2003,Lippert1997,Lippert1999,Fuesti-Molnar2005},
and the pseudo-spectral method and its variants\cite{Murphy2000,Neese2009}, can be considered to fall
under the general rubric of density fitting (DF) methods.
Density fitting can be used both when computing individual ERIs, as well as in intermediate formation.
The basic idea is to approximate the two-center atomic orbital pair density in Eq.~(\ref{eq:eri}) with
an expansion in auxiliary functions, the fitting basis. The approximate density $\rho'$ is obtained by minimizing its
distance to the reference two-center density $\rho$ with respect to a metric $g(\mathbf{r}_1,\mathbf{r}_2)$ (such as the Coulomb metric $r_{12}^{-1}$ or overlap metric $\delta(\mathbf{r}_1-\mathbf{r}_2)$)
\begin{equation*}
\min_{\rho'}  \iint[\rho(\mathbf{r}_1) - \rho'(\mathbf{r}_1)] g(\mathbf{r}_1,\mathbf{r}_2)
  [\rho(\mathbf{r}_2) - \rho'(\mathbf{r}_2)] d\mathbf{r}_1 d\mathbf{r}_2 \label{eq:fitrho}
\end{equation*}
By choosing different metrics and fitting bases, one recovers the different schemes mentioned above.
However, the most common version of DF uses a Gaussian fitting basis, in conjunction with the Coulomb metric. 
We will refer to this standard combination of fitting basis and Coulomb metric as Gaussian density fitting (GDF).
Gaussian density fitting is available in almost all the major quantum chemistry
packages today~\cite{Feyereisen1993,Komornicki1993,Vahtras1993,Eichkorn1995,Dunlap2000,Weigend1998,Sierka2003,Manby2001,Werner2003,Sodt2006,Werner2006,Jung2005}. 

In this work, we extend the Gaussian DF methodology to a {\it mixed basis} density fitting (MDF). This creates
an efficient DF framework well suited to the all-electron modeling of periodic systems. 
The basic idea in MDF is to use a mixed auxiliary basis of Gaussians $\chi_Q(\mathbf{r})$ and 
plane-waves (PW), expanding the density as
\begin{equation}
  \rho(\mathbf{r}) = \sum_Q \chi_Q(\mathbf{r}) d_{Q}
  + \sum_{\mathbf{G}} e^{i\mathbf{G}\cdot\mathbf{r}} c_\mathbf{G}
  \label{eq:mixedexpansion}
\end{equation}
The mixed-basis representation allows the representation of compact densities through the Gaussian functions $\chi_Q(\mathbf{r})$,
while offering systematic convergence for  smooth densities through the PWs.
These two properties  address the challenges of Coulomb evaluation in all-electron periodic calculations,
where contributions from both the core and diffuse interstitial densities must be efficiently computed. 
Further, the
use of a PW representation provides a natural way to handle the Coulomb divergence that appears in periodic settings.
Although such all-electron calculations can be expected to be more expensive than pseudo-potential calculations,
they allow us to carry out computations free of pseudo-potential error.

There are related works in the literature. These include the Gaussian and (augmented) plane-wave
formalism by Parrinello and coworkers~\cite{Lippert1997,krack2000all,vandevondele2005quickstep}, and the
Fourier transform Coulomb method of F\"{u}sti-Moln\'{a}r and Pulay~\cite{Fuesti-Molnar2002,Fuesti-Molnar2002a}.
In both of these,  Gaussian basis sets are used to expand the orbitals, and the density matrix contributions
of Gaussians with large exponents (compact Gaussians) and small exponents (smooth Gaussians) are separated.
The Coulomb potential and energy contributions of the smooth Gaussians are evaluated by PW density fitting using the FFT, while
the compact Gaussian ERIs are evaluated explicitly. Thus, unlike in our mixed density fitting, Gaussian density fitting is not used
at all. Further, both works are concerned with optimizing the evaluation of the Coulomb potential and energy
only, rather than the more general ERI kernel, as used in the computation of exchange and in many-body methods.
Some other differences include the manner in which compact and smooth densities are partitioned, 
as well as our use of analytical Fourier transforms to achieve higher accuracy than the FFT with the same number of PW's.
The impact of these choices will become apparent in the benchmark applications discussed below.

The rest of this manuscript describes in detail the implementation of the mixed density fitting scheme
and its benchmarking.
In sections \ref{sec:mdf} and \ref{sec:lindep} we present the formulae to
compute the 4-index ERIs in terms of the MDF mixed Gaussian and PW fitting basis.
The procedure to carry out GDF in a periodic system, which serves as a comparison for MDF, is discussed in section \ref{sec:df}
using some formulae developed in the MDF framework.
The MDF scheme is benchmarked for
 the all-electron Coulomb, exchange, and total
 energy at the Hartree-Fock level, and the all-electron band structure
 at the density functional level, for some simple crystals in section \ref{sec:examples}.
Our conclusions are presented in section \ref{sec:conclusions}.

\section{Theory}
\subsection{Mixed density fitting method for periodic systems}
\label{sec:mdf}
In an $N$-cell crystalline system, the AO functions $\phi_\mu(\mathbf{r})$ are
translational-symmetry-adapted linear combinations of Gaussian atomic
orbitals $\mu(\mathbf{r})$~\cite{Dovesi2005}
\begin{align}
  \phi_\mu(\mathbf{r})
  = \sum_\mathbf{T} e^{i\mathbf{k}_{\mu}\cdot\mathbf{T}}
  \mu(\mathbf{r}-\mathbf{T})
\end{align}
where $\mathbf{T}$ is a translation vector and $\mathbf{k}_{\mu}$ is a
crystal momentum vector.
In the mixed density fitting scheme, the AO products $\rho_{\mu\nu}(\mathbf{r})$ are
approximated by an expansion of periodic Gaussian fitting functions plus plane-wave
functions
\begin{align}
  \rho_{\mu\nu}(\mathbf{r}) = \phi_\mu^*(\mathbf{r})\phi_\nu(\mathbf{r})
  = \sum_Q \phi_Q^{\mathbf{k}_{\mu\nu}}(\mathbf{r}) d_{Q,\mu\nu}
  + \sum_{\mathbf{G}+\mathbf{k}_{\mu\nu}\neq 0} \frac{ e^{i(\mathbf{G}+\mathbf{k}_{\mu\nu})\cdot \mathbf{r}}}{\sqrt{N\Omega}} c_{\mathbf{G},\mu\nu}
  + \bar{\rho}_{\mu\nu},
  \label{eq:pbcexpansion}
\end{align}
where $\mathbf{k}_{\mu\nu} =-\mathbf{k}_\mu + \mathbf{k}_\nu$.
$N\Omega$ represents the total volume of the computational crystal and $\Omega$ is the
volume of the unit cell.
The fitting function $\phi_Q^\mathbf{k}(\mathbf{r})$ is defined as
\begin{gather}
  \phi_Q^{\mathbf{k}}(\mathbf{r})
  = \chi_Q^{\mathbf{k}}(\mathbf{r}) - \xi_Q^{\mathbf{k}}(\mathbf{r})
  = \frac{1}{\sqrt{N}}\sum_\mathbf{T} e^{i\mathbf{k}\cdot\mathbf{T}}
  [\chi_Q(\mathbf{r}-\mathbf{T})-\xi_Q(\mathbf{r}-\mathbf{T})],
  \label{eq:pbcgaussian}
  \\
  \chi_Q^{\mathbf{k}}(\mathbf{r})
  = \frac{1}{\sqrt{N}}\sum_\mathbf{T} e^{i\mathbf{k}\cdot\mathbf{T}} \chi_Q(\mathbf{r}-\mathbf{T}),
  \\
  \xi_Q^{\mathbf{k}}(\mathbf{r})
  = \frac{1}{\sqrt{N}}\sum_\mathbf{T} e^{i\mathbf{k}\cdot\mathbf{T}} \xi_Q(\mathbf{r}-\mathbf{T}),
\end{gather}
Here, $\chi_Q(\mathbf{r})$ is a compact Gaussian fitting function, and $\xi_Q(\mathbf{r})$
is a smooth Gaussian fitting function which is subtracted from it to ensure that the  fitting basis
functions carry zero net charge and zero multipoles.  For example, for a $p$-type
auxiliary function, we require that the dipole integral vanishes
\begin{equation}
  \int \mathbf{r}[\chi_p(\mathbf{r}) - \xi_p(\mathbf{r})] d\mathbf{r} = 0.
\end{equation}
 The Coulomb potential of a zero-charge and zero-multipole density decays exponentially in real space, and this
allows us to compute the Coulomb integrals of the Gaussian fitting functions using lattice summation.
The compensating function $\xi_Q(\mathbf{r})$ does not hold any other physical significance,
but should be chosen to be smooth so that its contributions can be efficiently compensated for in the PW expansion. 
Given a real space lattice sum truncation distance, the smoothness of $\xi_Q(\mathbf{r})$ can be optimized in
the same manner as is done in the optimization of the Ewald parameter\cite{Kolafa1992,gibbon2002}.
Because the charge is excluded from the Gaussian fitting basis, we handle it 
as part of the PW expansion, and this is the
last term in Eq. \eqref{eq:pbcexpansion} (corresponding to
$\mathbf{G} = 0$ and $\mathbf{k}_\mu=\mathbf{k}_\nu$) 
\begin{align}
  \bar{\rho}_{\mu\nu}
  &=\frac{1}{N\Omega}\int \phi_\mu^*(\mathbf{r}) \phi_\nu(\mathbf{r}) d\mathbf{r}
  = \frac{S_{\mu\nu}}{\Omega}
\end{align}
where
\begin{align}
  S_{\mu\nu} = \sum_\mathbf{T} e^{i\mathbf{k}_\nu\cdot \mathbf{T}}
  \int \mu^*(\mathbf{r}) \nu(\mathbf{r}-\mathbf{T}) d\mathbf{r}
\end{align}
is the AO overlap integral (per unit cell).

The fitting coefficients are obtained by minimizing the density fitting error in the Coulomb metric. This leads to
a linear equation for the coefficients $d_{Q,\mu\nu}$ and $c_{\mathbf{G},\mu\nu}$
\begin{align}
  \begin{pmatrix}
    (\phi_P^{-\mathbf{k}_{\mu\nu}}|\phi_Q^{\mathbf{k}_{\mu\nu}}) &
    \frac{4\pi \rho_P(-\mathbf{G}-\mathbf{k}_{\mu\nu})}{\sqrt{\Omega}|\mathbf{G}+\mathbf{k}_{\mu\nu}|^2} \\
    \frac{4\pi \rho_Q( \mathbf{G}+\mathbf{k}_{\mu\nu})}{\sqrt{\Omega}|\mathbf{G}+\mathbf{k}_{\mu\nu}|^2} &
    \frac{4\pi}{|\mathbf{G}+\mathbf{k}_{\mu\nu}|^2}
  \end{pmatrix} 
  \begin{pmatrix}
    d_{Q,\mu\nu} \\
    c_{\mathbf{G},\mu\nu}
  \end{pmatrix}
  =
  \begin{pmatrix}
    \sqrt{N}\left[(\phi_P^{-\mathbf{k}_{\mu\nu}}|\phi_\mu\phi_\nu) -
    \bar{V}_P^{-\mathbf{k}_{\mu\nu}}\bar{\rho}_{\mu\nu}\right] \\
    \frac{4\pi\rho_{\mu\nu}(\mathbf{G}+\mathbf{k}_{\mu\nu})}{\sqrt{\Omega}|\mathbf{G}+\mathbf{k}_{\mu\nu}|^2}
  \end{pmatrix}
  \label{eq:mdfeq}
\end{align}
where the integrals (derived in Appendix~\ref{sec:AppendixA}) are
\begin{align}
  (\phi_P^{-\mathbf{k}}|\phi_Q^{\mathbf{k}})
  &=\sum_\mathbf{T} e^{i\mathbf{k}\cdot\mathbf{T}}
  \int [\chi_P(\mathbf{r}_1)-\xi_P(\mathbf{r}_1)] \frac{1}{r_{12}}
  [\chi_Q(\mathbf{r}_2-\mathbf{T})-\xi_Q(\mathbf{r}_2-\mathbf{T})] d\mathbf{r}_1 d\mathbf{r}_2
  \label{eq:int2c2e}
  \\
  \rho_Q(\mathbf{G}+\mathbf{k}) 
  &=\int e^{-i(\mathbf{G}+\mathbf{k})\cdot\mathbf{r}}
  [\chi_Q(\mathbf{r})-\xi_Q(\mathbf{r})] d\mathbf{r}
  \label{eq:ftaux}
  \\
  \bar{V}_P^{\mathbf{k}}
  &=
  \begin{cases}
    \frac{\pi}{\alpha_{P_\xi}} - \frac{\pi}{\alpha_{P_\chi}}
      & \mathbf{k} = 0 \text{ and } \chi_P \in \text{s-type GTOs} \\
    0 & \text{otherwise}
  \end{cases}
  \label{eq:vbar}
  \\
  (\phi_P^{-\mathbf{k}_{\mu\nu}}|\phi_\mu\phi_\nu)
  &=\sum_{\mathbf{T}_\mu\mathbf{T}_\nu} e^{i\mathbf{k}_\nu\cdot\mathbf{T}_\nu-i\mathbf{k}_\mu\cdot\mathbf{T}_\mu}
  \int [\chi_P(\mathbf{r}_1)-\xi_P(\mathbf{r}_1)] \frac{1}{r_{12}}
  \mu^*(\mathbf{r}_2-\mathbf{T}_\mu)\nu(\mathbf{r}_2-\mathbf{T}_\nu) d\mathbf{r}_1 d\mathbf{r}_2
  \label{eq:int3c2e}
  \\
  \rho_{\mu\nu}(\mathbf{G}+\mathbf{k}_{\mu\nu})
  &=\sum_{\mathbf{T}} e^{i\mathbf{k}_\nu\cdot\mathbf{T}}
  \int e^{-i(\mathbf{G}+\mathbf{k}_{\mu\nu})\cdot\mathbf{r}}
  \mu^*(\mathbf{r})\nu(\mathbf{r}-\mathbf{T}) d\mathbf{r}
  \label{eq:ftaoao}
\end{align}
In Eq.~\eqref{eq:vbar}, $\alpha_{P_\chi}$ and $\alpha_{P_\xi}$ are the exponents of
the Gaussian functions $\chi_P(\mathbf{r})$ and $\xi_P(\mathbf{r})$.
In the above integrals, computing the three center integral \eqref{eq:int3c2e}
is  demanding due to the double lattice sum, with a cost
of $O(n^2 m N_c^2)$ where $n$ is the number of AOs, $m$ is the number of
auxiliary Gaussian functions and $N_c$ is the number of images in the lattice
summation.

Eqs.~\eqref{eq:ftaux} and \eqref{eq:ftaoao} involve the Fourier
transforms of the fitting Gaussians and AO products.
While one can approximate these integrals using a
discrete fast Fourier transform (FFT), this is only practical
if the Gaussians involved are not very steep, as for example,
in pseudo-potential calculations; otherwise
prohibitively large Fourier grids are necessary (see section~\ref{sec:examples}).
Alternatively, the integrals can be calculated analytically.
The formulae for the analytical
Fourier transforms are documented in Appendix \ref{sec:AppendixB}.
The leading computational cost is for the AO products which has 
a formal scaling of $O(n^2N_GN_c)$ where $N_G$ is the number of PWs. Although
there is only one factor of $N_c$ (compared to the three center Gaussian integrals)
the analytical Fourier transforms also become expensive 
for a large number of PWs. However, as long as
the Gaussian fitting functions of core and valence characters are appropriately tuned, it is not
difficult to require only a modest number of PWs in the MDF expansion
of the smooth part of the density. 
The analytical Fourier transform technique can be used in both pseudo-potential and
all-electron calculations. As the PWs in this approach are strictly used
only to represent the density and not to numerically sample the Gaussians,
one can use fewer PWs with the analytical Fourier transform than in a typical FFT-driven calculation.




Finally, when defining ERIs in a periodic system, we remove the net charge
of the AO product to avoid the divergent Coulomb contribution, corresponding to removing the $\mathbf{G}=0$
singularity when $\mathbf{k}_{\mu\nu}=0$~\cite{McClain2017}. (The $\mathbf{G}=0$ electronic contribution, which
only depends on the number of electrons in the unit cell, is appropriately
handled together with the electron-nuclear and nuclear-nuclear $\mathbf{G}=0$ contributions, yielding an additive constant
to the total energy~\cite{McClain2017}.)
Using the quantities defined in the MDF expansion,
the periodic ERI (here, and in the following text, per unit cell) is assembled as
\begin{align}
  (\mu\nu|\kappa\lambda)
  &=\frac{1}{N}\int \frac{\left[\phi_\mu(\mathbf{r}_1)\phi_\nu(\mathbf{r}_1)-\bar{\rho}_{\mu\nu}\right]
  \left[\phi_\kappa(\mathbf{r}_2)\phi_\lambda(\mathbf{r}_2)-\bar{\rho}_{\kappa\lambda}\right]}{r_{12}}
  d\mathbf{r}_1d\mathbf{r}_2
  \notag\\
  &=\sum_Q \frac{d_{Q,\mu\nu}}{\sqrt{N}}(\phi_Q^{\mathbf{k}_{\mu\nu}}|\phi_\kappa\phi_\lambda)
  - \bar{\rho}_{\kappa\lambda}\sum_Q \frac{d_{Q,\mu\nu}}{\sqrt{N}}\bar{V}_Q^{\mathbf{k}_{\mu\nu}}
  + \sum_{\mathbf{G}+\mathbf{k}_{\mu\nu}\neq 0} c_{\mathbf{G},\mu\nu}
  \rho_{\kappa\lambda}(-\mathbf{G}+\mathbf{k}_{\kappa\lambda})
\label{eq:pbceri}
\end{align}
In the ERI expression, crystal momentum conservation is
used
\begin{equation}
  (-\mathbf{k}_\mu + \mathbf{k}_\nu - \mathbf{k}_\kappa + \mathbf{k}_\lambda)
  \cdot\mathbf{a} = 2n\pi
\end{equation}
where $\mathbf{a}$ is the lattice vector.

\subsection{Linear dependence in the MDF fitting basis}
\label{sec:lindep}
In the mixed fitting basis, the periodic Gaussian functions and PWs may become linearly
dependent with respect to each other as each subset becomes increasingly complete.  In
practice, this  causes numerical instabilities when solving the linear equation
in the form \eqref{eq:mdfeq} directly.  To remove the linear dependencies, we orthogonalize
the fitting functions with respect to the Coulomb metric through the transformation
\begin{gather}
  \begin{pmatrix}
    \phi_Q^k(\mathbf{r}) & \frac{e^{i(\mathbf{G}+\mathbf{k})\cdot\mathbf{r}}}{\sqrt{N\Omega}}
  \end{pmatrix}
  \begin{pmatrix}
    \mathbf{t}^\mathbf{k} & 0 \\
    -\frac{\rho_Q(\mathbf{G}+\mathbf{k})}{\sqrt{\Omega}}\mathbf{t}^\mathbf{k} &
    \frac{|\mathbf{G}+\mathbf{k}|}{2\sqrt{\pi}}
  \end{pmatrix}
  \label{eq:lindep}
\end{gather}
where the rectangular matrix $\mathbf{t}^{\mathbf{k}}$ is the transformation to diagonalize
the dressed Coulomb matrix of the Gaussian fitting functions
\begin{gather}
  \tilde{J}_{PQ}^\mathbf{k} = (\phi_P^{-\mathbf{k}}|\phi_Q^{\mathbf{k}})
  - \sum_{\mathbf{G}+\mathbf{k}\neq 0}\frac{4\pi\rho_P(-\mathbf{G}-\mathbf{k})\rho_Q(\mathbf{G}+\mathbf{k})}
  {\Omega|\mathbf{G}+\mathbf{k}|^2}
  \label{eq:coulmetric}
  \\
  \mathbf{t}^{\mathbf{k}\dagger} \tilde{\mathbf{J}}^\mathbf{k}
  \mathbf{t}^\mathbf{k} = \mathbf{1}
\end{gather}
Although different choices can be made to remove  linear dependencies, different schemes do
not  share the same numerical stability.
We used the transformation \eqref{eq:lindep} because it does not mix Gaussian
functions into the PWs.
An advantage of the PW basis is that the Coulomb operator is diagonal in the PW
representation.  
Manipulating the basis orthogonalization in this diagonal representation is
straightforward and numerically stable, leading to the normalization factor
$|\mathbf{G}+\mathbf{k}|/2\sqrt{\pi}$ in Eq. \eqref{eq:lindep}.
Projecting the PWs out of the Gaussian functions in \eqref{eq:coulmetric} leads to a highly singular
matrix.  To remove the linear dependence of the Gaussian
functions, we diagonalize this singular matrix and remove the eigenvectors
associated with small eigenvalues below a threshold.
The effect of the linear dependence
threshold on the stability of the results is tested in Section \ref{sec:examples}.
In our program, we use a default threshold of $10^{-7}$.

Applying transformation \eqref{eq:lindep} to the
linear equation \eqref{eq:mdfeq} followed by removal of small eigenvalues allows us to stably determine
 the density fitting coefficients. With respect
to the transformed fitting functions, we can define a new MDF expression for the ERIs in \eqref{eq:pbceri} 
\begin{gather}
  (\mu\nu|\kappa\lambda)
  =\sum_i L_{i,\mu\nu}L_{i,\kappa\lambda}
  +\sum_{\mathbf{G}+\mathbf{k}_{\mu\nu} \neq 0}
  \frac{4\pi\rho_{\mu\nu}(\mathbf{G}+\mathbf{k}_{\mu\nu})\rho_{\kappa\lambda}(-\mathbf{G}+\mathbf{k}_{\kappa\lambda})}
  {\Omega|\mathbf{G}+\mathbf{k}_{\mu\nu}|^2}
  \label{eq:pbceri1}
  \\
  L_{i,\mu\nu} = \sum_P t_{Pi}^{\mathbf{k}_{\mu\nu} *}
  \left[(\phi_P^{-\mathbf{k}_{\mu\nu}}|\phi_\mu\phi_\nu) -
  \bar{V}_P^{-\mathbf{k}_{\mu\nu}}\bar{\rho}_{\mu\nu} -
  \sum_{\mathbf{G}+\mathbf{k}_{\mu\nu}\neq 0}\frac{4\pi\rho_P(-\mathbf{G}-\mathbf{k}_{\mu\nu})\rho_{\mu\nu}(\mathbf{G}+\mathbf{k}_{\mu\nu})}
  {\Omega|\mathbf{G}+\mathbf{k}_{\mu\nu}|^2}\right]
  \label{eq:Lpij}
\end{gather}

\subsection{Gaussian density fitting for periodic systems}
\label{sec:df}

In the current work, we will benchmark mixed density fitting against standard Gaussian density fitting.
We first describe how GDF may be efficiently implemented in periodic systems
using some of the results introduced above for MDF.
In the periodic setting, the AO products in the standard GDF method are expanded in a set of periodic Gaussian
fitting functions
\begin{gather}
  \rho_{\mu\nu}
  = \sum_Q \chi_Q^{\mathbf{k}_{\mu\nu}}(\mathbf{r}) d_{Q,\mu\nu}.
\end{gather}
The Coulomb metric when used directly in the periodic setting diverges.
Thus, we
exclude the net charge of the AO products in the fitting expansion
\begin{gather}
  \rho_{\mu\nu}(\mathbf{r}) - \bar{\rho}_{\mu\nu}
  = \sum_Q \left(\chi_Q^{\mathbf{k}_{\mu\nu}}(\mathbf{r})
  - \bar{\chi}_Q^{\mathbf{k}_{\mu\nu}}\right) d_{Q,\mu\nu}.
  \\
  \bar{\chi}_Q^{\mathbf{k}}
  =
  \begin{cases}
    \frac{\sqrt{N}}{\Omega} & \mathbf{k} = 0 \text{ and } \chi_Q(\mathbf{r}) \in s\text{-type GTOs} \\
    0 & \text{otherwise}
  \end{cases}
\end{gather}
The two-electron integrals can then be formulated in terms of the GDF quantities as
\begin{gather}
  (\mu\nu|\kappa\lambda)
  =\sum_i \mathcal{V}_{P,\mu\nu}(\mathcal{J}^{-1})_{PQ}\mathcal{V}_{Q,\kappa\lambda}
  \\
  \mathcal{V}_{P,\mu\nu}
  =\frac{1}{\sqrt{N}}(\chi_P^{\mathbf{k}_{\nu\mu}}-\bar{\chi}_P^{\mathbf{k}_{\nu\mu}}
  | \phi_\mu\phi_\nu - \bar{\rho}_{\mu\nu})
  \label{eq:df3c}
  \\
  \mathcal{J}_{PQ}
  =(\chi_P^{\mathbf{k}_{\nu\mu}}-\bar{\chi}_P^{\mathbf{k}_{\nu\mu}}
  | \chi_Q^{\mathbf{k}_{\mu\nu}}-\bar{\chi}_Q^{\mathbf{k}_{\mu\nu}})
  \label{eq:df2c}
\end{gather}
The two-center and three-center Coulomb integrals represent Coulomb
interactions between chargeless density distributions and thus
are not divergent in the real space lattice summation, however the
  convergence may be very slow or even conditional on the summation order.
To accelerate the lattice summation, we can insert a compensating
function $\xi^\mathbf{k}(\mathbf{r})$ in the density fitting
expansion that removes higher multipoles of $\chi_Q^{\mathbf{k}}(\mathbf{r})$ as in the MDF procedure, 
\begin{align}
  \rho_{\mu\nu}(\mathbf{r}) - \bar{\rho}_{\mu\nu}
  &=\sum_Q \left(\chi_Q^{\mathbf{k}_{\mu\nu}}(\mathbf{r})
  - \xi_Q^{\mathbf{k}_{\mu\nu}}(\mathbf{r})
  + \xi_Q^{\mathbf{k}_{\mu\nu}}(\mathbf{r})
  - \bar{\chi}_Q^{\mathbf{k}_{\mu\nu}}\right) d_{Q,\mu\nu}
  \notag\\
  &=\sum_Q \phi_Q^{\mathbf{k}_{\mu\nu}}(\mathbf{r}) d_{Q,\mu\nu}
  + \sum_Q \left(\xi_Q^{\mathbf{k}_{\mu\nu}}(\mathbf{r})
  - \bar{\chi}_Q^{\mathbf{k}_{\mu\nu}}\right) d_{Q,\mu\nu},
\end{align}
This allows us to efficiently compute the two-center and  three-center integrals in a two-step scheme:
first we evaluate the integrals involving $\phi_Q(\mathbf{r})$ using real space lattice
summation; then the remaining contributions are evaluated using a PW expansion.  With this scheme,
the integrals \eqref{eq:df3c} and \eqref{eq:df2c} are obtained as
\begin{align}
  \mathcal{V}_{P,\mu\nu}
  &=(\phi_P^{-\mathbf{k}_{\mu\nu}}|\phi_\mu\phi_\nu)
  - \bar{V}_P^{-\mathbf{k}_{\mu\nu}}\bar{\rho}_{\mu\nu}
  +\sum_{\mathbf{G}+\mathbf{k}_{\mu\nu}\neq 0}
  \frac{4\pi\rho_{\xi_P}(-\mathbf{G}-\mathbf{k}_{\mu\nu})
  \rho_{\mu\nu}(\mathbf{G}+\mathbf{k}_{\mu\nu})}
  {\Omega|\mathbf{G}+\mathbf{k}_{\mu\nu}|^2}
  \\
  \mathcal{J}_{PQ}
  &=(\phi_P^{-\mathbf{k}_{\mu\nu}} | \phi_Q^{ \mathbf{k}_{\mu\nu}})
  + \sum_{\mathbf{G}+\mathbf{k}_{\mu\nu}\neq 0}
  \frac{4\pi\rho_{\xi_P}(-\mathbf{G}-\mathbf{k}_{\mu\nu}) \rho_Q(\mathbf{G}+\mathbf{k}_{\mu\nu})}
  {\Omega|\mathbf{G}+\mathbf{k}_{\mu\nu}|^2}
  \notag\\
  &+\sum_{\mathbf{G}+\mathbf{k}_{\mu\nu}\neq 0}
  \frac{4\pi\rho_P(-\mathbf{G}-\mathbf{k}_{\mu\nu}) \rho_{\xi_Q}(\mathbf{G}+\mathbf{k}_{\mu\nu})}
  {\Omega|\mathbf{G}+\mathbf{k}_{\mu\nu}|^2}
  + \sum_{\mathbf{G}+\mathbf{k}_{\mu\nu}\neq 0}
  \frac{4\pi\rho_{\xi_P}(-\mathbf{G}-\mathbf{k}_{\mu\nu}) \rho_{\xi_Q}(\mathbf{G}+\mathbf{k}_{\mu\nu})}
  {\Omega|\mathbf{G}+\mathbf{k}_{\mu\nu}|^2}
\end{align}
where
\begin{equation}
  \rho_{\xi_P}(\mathbf{G}+\mathbf{k})
  = \int e^{-i(\mathbf{G}+\mathbf{k})\cdot \mathbf{r}} \xi_P(\mathbf{r}) d\mathbf{r}
\end{equation}
Note that in the GDF calculations, {\it we always use sufficient
  number of PWs to completely converge the PW representation of
  the compensating Gaussian}. This ensures that the GDF calculations are a measure
purely of the quality of the original Gaussian density fitting basis.

\section{Benchmarking MDF}
\label{sec:examples}

We have implemented the MDF method as described above in our electronic structure program package PySCF~\cite{pyscf}.
To test the accuracy of the MDF method, we first computed the $\Gamma$-point Hartree-Fock Coulomb
($E_J$) and exchange energies ($E_K$) for the hydrogen crystal (cubic unit cell in Fd$\bar{3}$m symmetry,
lattice parameter $a$ = 3.567 \AA, cc-pVDZ basis).
The MDF Gaussian fitting basis was the even tempered basis (ETB) $10s6p2d$ (see Table \ref{tab:auxbasis}).
The compensating Gaussians (see Eq. \eqref{eq:pbcgaussian}) were chosen to have exponent 0.2.
The PW basis was constructed from a uniform reciprocal grid.
The real-space lattice summation was truncated at a distance of 9.2 \AA.
This ensured that both the AO basis and auxiliary
Gaussian basis lattice sums were fully converged.

\begin{figure}[htp]
  \begin{center}
\begin{subfigure}[b]{0.5\textwidth}
  \centering
  \includegraphics[width=\textwidth]{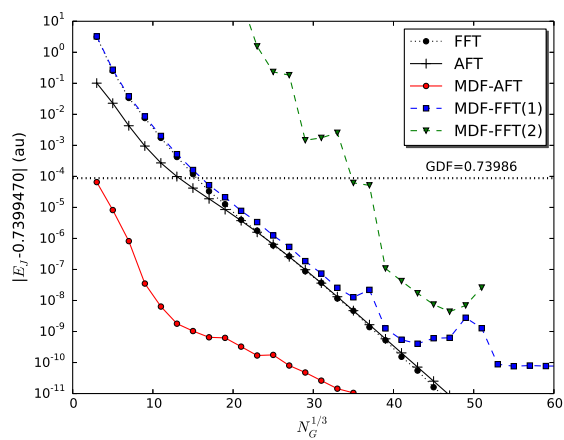}
  \caption{Coulomb energy for the H crystal}
  \label{fig:h8:a}
\end{subfigure}%
\begin{subfigure}[b]{0.5\textwidth}
  \centering
  \includegraphics[width=\textwidth]{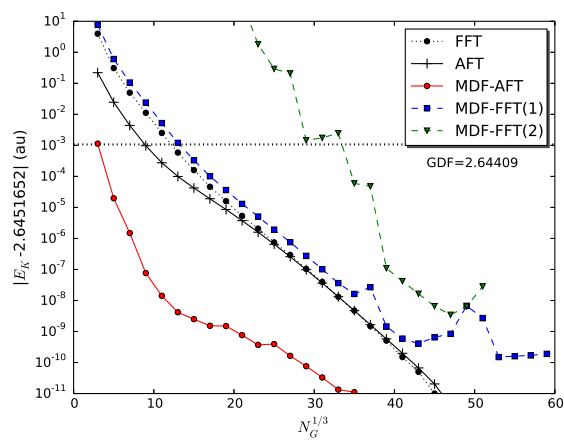}
  \caption{Exchange energy for the H crystal}
  \label{fig:h8:b}
\end{subfigure}%
  \end{center}
  \caption{Coulomb and exchange interactions per unit cell for the H crystal ($\Gamma$ point).
  In the H crystal, the reference energies are computed using a pure PW fitting basis
  and FFT computation of all terms with $N_{G}^{1/3}=101$ grid points.
  The ETB basis $10s6p2d$ is employed for MDF and GDF. }
  \label{fig:h8}
\end{figure}

We compare the different kinds of density fitting in 
Figures \ref{fig:h8:a} and \ref{fig:h8:b}.
Using pure GDF and the large even tempered fitting basis, we can fit $E_J$ to roughly 0.1 m$E_\mathrm{h}$ accuracy
and $E_K$ to roughly 1 m$E_\mathrm{h}$ accuracy. 
Note that the H atom cc-pVDZ basis does not contain any steep Gaussian functions.
This means it is also practical in this system to use only PWs as the fitting functions.
We show the results of PW density fitting (labelled FFT) where the PW coefficients and contributions are
determined by FFT. The PW density fitting converges the Coulomb and exchange energy very systematically
as a function of the number of PWs. This demonstrates the strength of including PWs in the fitting basis, and in fact
we use the systematic convergence to estimate the reference Coulomb and exchange energies.
Finally, we observe the effect of using both Gaussians and PWs in the MDF expansion (labelled MDF-AFT).
We see that  introducing the Gaussian fitting basis leads to
improved convergence relative to the pure PW expansion. The MDF expansion is 4-5
orders of magnitude more accurate than the pure PW expansion with the same number of PWs.
The accuracy of GDF itself is close to the accuracy of MDF with a minimal PW basis (27 PWs, 1 grid point per direction).
Since the difference between the Gaussian fitting basis in GDF and MDF 
is the set of compensating Gaussian functions in MDF, this 
reflects the fact that the compensating Gaussians used in MDF are here well represented by a small number of PWs.
Further, adding a modest number of PWs in MDF significantly improves the accuracy over
pure GDF, for example, 729 PWs (9 per axis) reduces the fitting error to 0.1
$\mu E_\mathrm{h}$.

In the MDF-AFT results, we used analytical Fourier transforms for all
PW-related integrals in the MDF method.
As discussed in the methods section, it is also possible to use
the discrete FFT to compute these integrals, although additional errors
are expected.
Note that there are three equations \eqref{eq:coulmetric}, \eqref{eq:pbceri1} and
\eqref{eq:Lpij} that involve quantities in reciprocal space. 
FFT cannot be used to obtain the reciprocal space densities in Eq. \eqref{eq:coulmetric} because
the numerical FFT destroys the positive definiteness of the metric.
We tested the use of the FFT integrals in the other two equations as follows:
(1) Using FFT reciprocal space quantities in the second term of Eq. \eqref{eq:pbceri1}, 
denoted MDF-FFT(1) in Figures \ref{fig:h8:a} and \ref{fig:h8:b}; (2)
Using FFT reciprocal space quantities in both Eqs. \eqref{eq:pbceri1} and \eqref{eq:Lpij}, denoted MDF-FFT(2).
To illustrate the density sampling error when using the FFT, we
also computed the PW related integrals using AFT, and the pure AFT results are also presented in 
Figures \ref{fig:h8:a} and \ref{fig:h8:b}.
MDF-FFT(1) gives a similar error to pure PW density fitting (using FFT for the PW related integrals)
because the errors from the FFT density sampling is larger than the corrections
introduced by the Gaussian fitting functions in MDF.
The error in MDF-FFT(2) is more severe, as the numerical errors introduced by the FFT
are compounded in Eqs. \eqref{eq:pbceri1} and \eqref{eq:Lpij}.
In either case, the use of the FFT to approximate the quantities involved in MDF clearly leads to unacceptable errors.

\begin{figure}[htp]
  \centering
  \includegraphics[width=\textwidth]{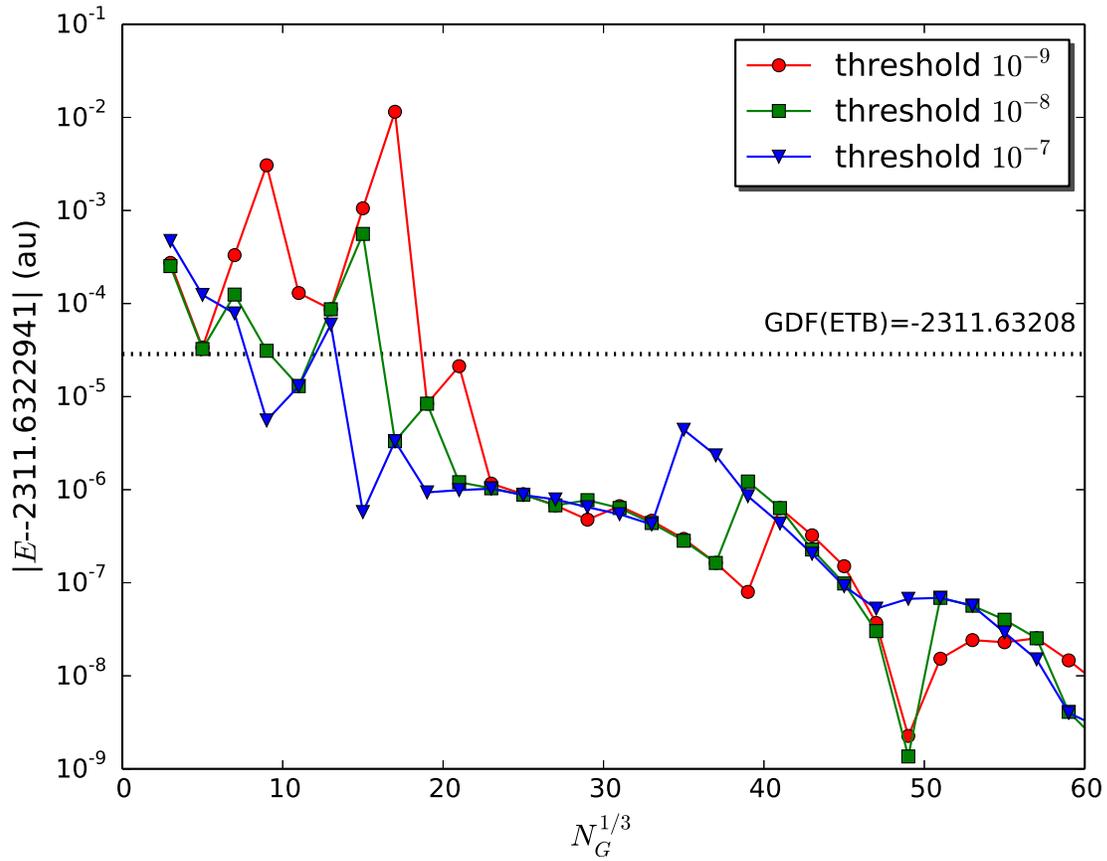}
  \caption{Hartree-Fock energy per unit cell for the Si crystal.
  The reference energies are computed using MDF with $N_{G}^{1/3}=81$ grid
  points.  The ETB basis is $20s16p13d7f2g$.  }
  \label{fig:si8hf}
\end{figure}

Figure \ref{fig:si8hf} shows the convergence of  $\Gamma$-point all-electron Hartree-Fock
energies for the silicon crystal
(Fd$\bar{3}$m symmetry, lattice parameter $a$ = 5.431 \AA, cc-pVDZ basis).
We used a large ETB fitting basis $20s16p13d7f2g$  (see Table
\ref{tab:auxbasis}).  The exponents of the compensating Gaussians were set to
0.2, and the real space lattice sums were truncated at 12 \AA. We use this system
to test the effect of the linear dependency threshold, and linear
thresholds of $10^{-9}$, $10^{-8}$ and $10^{-7}$ were used, keeping all 
other settings the same.

Because of the large Gaussian fitting basis, the GDF method achieves good
accuracy for the total energy. This is close to the accuracy of the MDF method with
125 PWs (corresponding to a PW energy cutoff of 20 eV).
In this system, the presence of core functions means that PWs alone
are insufficient to expand the densities; however, when used in conjunction with Gaussians, MDF systematically
improves beyond the GDF result.
However, we observe  that the linear dependency threshold in MDF can strongly
affect the accuracy of the HF energy.  
With exact arithmetic, a tighter linear dependence threshold, which retains more fitting
functions, should produce more accurate results.
However, for a small number of PWs, the thresholds $10^{-9}$ and $10^{-8}$ in
fact introduce large errors, due to the numerical instability associated with the
linear dependence between diffuse Gaussians and PWs.
Thus although more diffuse fitting functions are removed by the looser threshold,
the numerical problems are less severe and better accuracy is achieved.
Taking the MDF basis with 729 PWs (energy cutoff 80 eV) as an example,
a threshold of $10^{-9}$ retains 1471
fitting functions out of the original 1600 auxiliary Gaussian functions and gives
an error of 10 m$E_\mathrm{h}$, while a threshold of $10^{-7}$ retains 1312 linearly independent fitting
functions and gives an error of only 0.01 m$E_\mathrm{h}$.
However, when higher energy PWs are included in the MDF expansion, 
more diffuse Gaussians are removed by the threshold, and the
PW functions increasingly take over the role of expanding the diffuse density.
In this case, the numerical issues become relatively less serious and the different thresholds
produce similar convergence, since the linear dependence between the steep
Gaussian functions and the PWs is weak.  For example, with 24389 PWs (energy cutoff
1000 eV) in MDF, a threshold of $10^{-9}$ leads to 944 linearly independent auxiliary
Gaussian functions,  and a threshold of $10^{-7}$ leads to 888 functions.

\begin{figure}[htp]
  \centering
  \includegraphics[width=\textwidth]{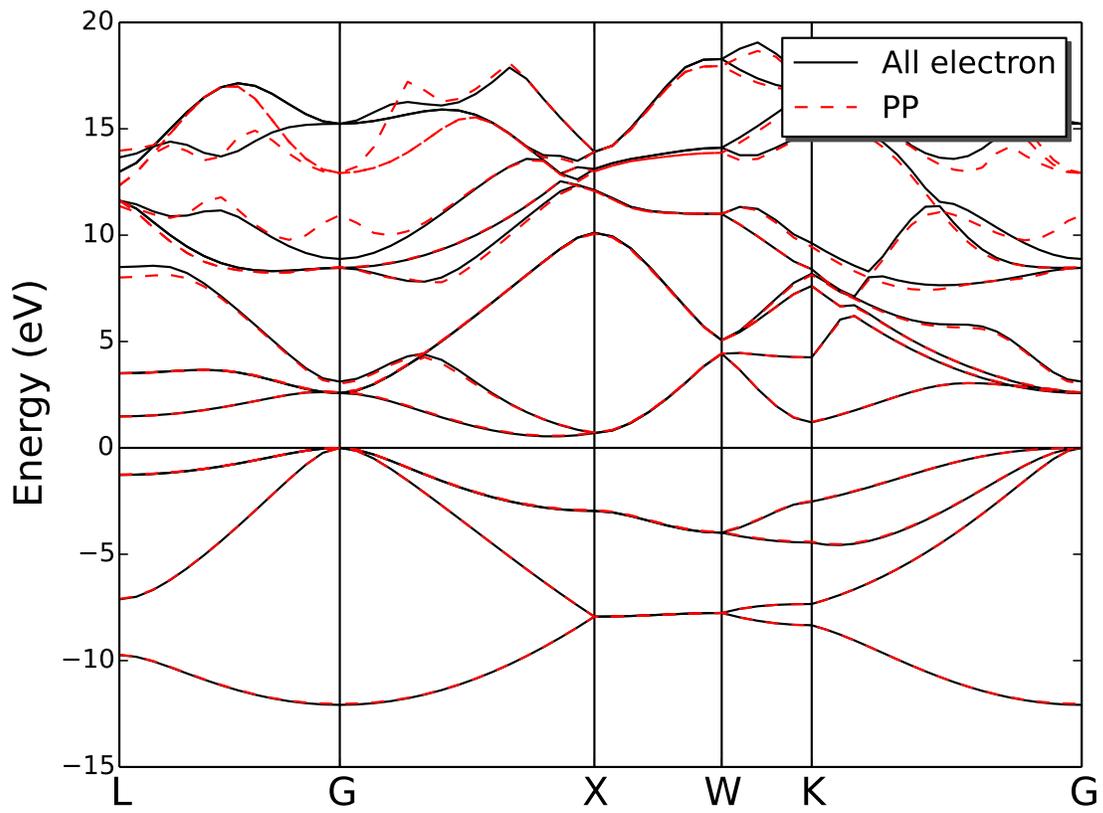}
  \caption{All-electron and pseudopotential LDA band structure of the Si crystal.}
  \label{fig:siband}
\end{figure}

As discussed in the introduction, an important motivation for all-electron calculations
enabled by  MDF is that they  allow us to assess  pseudo-potential error. We
now briefly examine the pseudo-potential error in the band structure of the silicon crystal.  Figure
\ref{fig:siband} presents the LDA bands computed within a pseudopotential (PP)
and an all-electron calculation using a $6 \times 6 \times 6$ $k$-point mesh with two
atoms per (primitive) unit cell.  In the PP calculation, we used the GTH
pseudopotentials\cite{Goedecker1996,Hartwigsen1998} that were optimized for the LDA functional and the
GTH DZVP basis, obtained from the CP2K\cite{WCMS-CP2K,vandevondele2005quickstep} software package.  The PP Coulomb
integrals were computed with 3375 PWs (energy cutoff 750 eV) using FFT.
In the all-electron calculation, we used the cc-pVDZ AO basis and a fitting
basis consisting of the ETB basis $20s16p13d7f2g$ and 1331 PWs (energy
cutoff 380 eV).  The valence bands and conduction bands agree very well between the two
types of calculations near the Fermi level, although quantitative discrepancies
appear further from the Fermi level. The band gap of the PP calculation is 0.72 eV while the
all electron calculation predicts a band gap of 0.69 eV.

Last, we briefly compare the computational cost of the MDF method 
in all-electron and pseudo-potential calculations to our earlier Gaussian orbital FFT-based pseudo-potential algorithm~\cite{pyscf,McClain2017}.
In the FFT-based DFT calculation, evaluating the Gaussian AO values on the real-space mesh
grid is the expensive operation with a formal scaling of $O(nN_GN_c)$.
As shown in Section \ref{sec:mdf}, the scaling of the AFT in the MDF integrals
is $n$ times higher than the scaling of AO evaluation.
In addition to the analytical Fourier transforms, the MDF method also requires the
three-index Gaussian integrals, and these are computationally demanding as well.
In the applications to the silicon crystal test system above using pseudopotentials, 
we found that the cost of the pseudo-potential MDF calculation was about an order
of magnitude higher than the pseudo-potential FFT-based calculation. This reflects
the fact that the Gaussian
fitting basis is not really required to represent purely smooth densities. However,
the strength of the MDF procedure is to enable all-electron calculations, and the
all-electron calculations using the pure FFT algorithm would be prohibitively (orders of magnitude more) expensive
than with the MDF implementation.

\section{Conclusions}
\label{sec:conclusions}
In this work, we presented a Gaussian and plane-wave mixed density fitting  (MDF) to
 compute electron repulsion integrals and  associated quantities
such as the Coulomb and exchange energies in periodic systems. 
Our algorithm possesses several new features, including 
the use of analytical Fourier transforms instead of the standard Fast Fourier
Transform to achieve high accuracy, and
an efficient transformation to remove linear dependencies between
Gaussians and PWs.

MDF allows for periodic calculations both with pseudo-potentials and with all electrons.
Compared to conventional GDF, the main advantage
of MDF is the ability to systematically converge to high accuracy through the PW
part of the expansion with a relatively weak dependence on the quality of
the Gaussian fitting basis, and without the need for diffuse Gaussian fitting functions. 
The main disadvantage of the technique
is the overhead incurred from handling the (relatively) large PW fitting basis.
This means that the MDF approach is unlikely to be the method of choice for low-accuracy, or pseudo-potential calculations.
However, for high accuracy all-electron calculations in large systems, MDF provides an efficient computational choice.
Further, it is possible to accelerate MDF calculations by exploiting the dual sparsity of the densities in real and reciprocal space.
These optimizations will be considered in our future work.


\begin{appendices}

\section{Analytical integrals for periodic Gaussian functions}
\label{sec:AppendixA}
The integrals we presented in Section \ref{sec:mdf} can be evaluated
analytically with real space lattice sums.
For a crystal-momentum-conserving AO basis ($\mathbf{k}_\mu=\mathbf{k}_\nu$),
the AO overlap integrals (per unit cell) can be computed as
\begin{align*}
  S_{\mu\nu}
  = \frac{1}{N}\langle \phi_\mu|\phi_\nu\rangle
  &= \frac{1}{N}\int
  \sum_{\mathbf{T}_\mu}e^{-i\mathbf{k}_\mu\cdot\mathbf{T}_\mu}\mu^*(\mathbf{r}-\mathbf{T}_\mu)
  \sum_{\mathbf{T}_\nu}e^{ i\mathbf{k}_\nu\cdot\mathbf{T}_\nu}\nu  (\mathbf{r}-\mathbf{T}_\nu) d\mathbf{r}
  \\
  &= \frac{1}{N}\int
  \sum_{\mathbf{T}_\mu}e^{-i\mathbf{k}_\mu\cdot\mathbf{T}_\mu}\mu^*(\mathbf{r}-\mathbf{T}_\mu)
  \sum_{\mathbf{T}_\nu}e^{ i\mathbf{k}_\nu\cdot(\mathbf{T}_\nu+\mathbf{T}_\nu)}
  \nu  (\mathbf{r}-\mathbf{T}_\mu-\mathbf{T}_\nu) d\mathbf{r}
  \\
  &= \frac{1}{N}\sum_{\mathbf{T}_\mu}e^{i(\mathbf{k}_\nu-\mathbf{k}_\mu)\cdot\mathbf{T}_\mu}
  \int \sum_{\mathbf{T}_\nu}\mu^*(\mathbf{r})
  e^{ i\mathbf{k}_\nu\cdot\mathbf{T}_\nu}\nu  (\mathbf{r}-\mathbf{T}_\nu) d\mathbf{r}
  \\
  &=\int \sum_{\mathbf{T}_\nu}\mu^*(\mathbf{r})
  e^{ i\mathbf{k}_\nu\cdot\mathbf{T}_\nu}\nu  (\mathbf{r}-\mathbf{T}_\nu) d\mathbf{r}
\end{align*}
A similar treatment can be used for the other Gaussian integrals and Fourier
transforms
\begin{align*}
  (\phi_P^{-\mathbf{k}}|\phi_Q^\mathbf{k})
  &=\int \phi_P^{-\mathbf{k}}(\mathbf{r}_1) \frac{1}{r_{12}}
  \phi_Q^{\mathbf{k}}(\mathbf{r}_2) d\mathbf{r}_1\mathbf{r}_2
  \\
  &=\sum_\mathbf{T} e^{i\mathbf{k}\cdot\mathbf{T}}
  \int [\chi_P(\mathbf{r}_1)-\xi_P(\mathbf{r}_1)] \frac{1}{r_{12}}
  [\chi_Q(\mathbf{r}_2-\mathbf{T})-\xi_Q(\mathbf{r}_2-\mathbf{T})] d\mathbf{r}_1 d\mathbf{r}_2
\end{align*}
\begin{align*}
  (\phi_P^{-\mathbf{k}_{\kappa\lambda}}|\phi_\mu\phi_\nu)
  &=\frac{1}{\sqrt{N}}\int \phi_P^{-\mathbf{k}_{\kappa\lambda}}(\mathbf{r}_1)
  \frac{1}{r_{12}} \phi_\mu^*(\mathbf{r}_2)\phi_\nu(\mathbf{r}_2) d\mathbf{r}_1d\mathbf{r}_2
  \\
  &=\sum_{\mathbf{T}_\mu\mathbf{T}_\nu} e^{i\mathbf{k}_\nu\cdot\mathbf{T}_\nu-i\mathbf{k}_\mu\cdot\mathbf{T}_\mu}
  \int [\chi_P(\mathbf{r}_1)-\xi_P(\mathbf{r}_1)] \frac{1}{r_{12}}
  \mu^*(\mathbf{r}_2-\mathbf{T}_\mu)\nu(\mathbf{r}_2-\mathbf{T}_\nu) d\mathbf{r}_1 d\mathbf{r}_2
\end{align*}
\begin{align*}
  \rho_P(\mathbf{G}+\mathbf{k})
  &=\frac{1}{\sqrt{N}}\int e^{-i(\mathbf{G}+\mathbf{k})\cdot\mathbf{r}}\phi_P^\mathbf{k}(\mathbf{r}) d\mathbf{r}
  \\
  &=\frac{1}{N}\int e^{-i(\mathbf{G}+\mathbf{k})\cdot\mathbf{r}}
  \sum_\mathbf{T}e^{i\mathbf{k}\cdot\mathbf{T}}
  [\chi_P(\mathbf{r}-\mathbf{T})-\xi_P(\mathbf{r}-\mathbf{T})] d\mathbf{r}
  \\
  &=\frac{1}{N}\int \sum_\mathbf{T}e^{-i(\mathbf{G}+\mathbf{k})\cdot(\mathbf{r}+\mathbf{T})}
  e^{i\mathbf{k}\cdot\mathbf{T}} [\chi_P(\mathbf{r})-\xi_P(\mathbf{r})] d\mathbf{r}
  \\
  &=\int e^{-i(\mathbf{G}+\mathbf{k})\cdot\mathbf{r}} [\chi_P(\mathbf{r})-\xi_P(\mathbf{r})] d\mathbf{r}
\end{align*}
\begin{align*}
  \rho_{\mu\nu}(\mathbf{G}+\mathbf{k}_{\mu\nu})
  &=\int e^{-i(\mathbf{G}+\mathbf{k}_{\mu\nu})}\phi_\mu^*(\mathbf{r}) \phi_\nu(\mathbf{r}) d\mathbf{r}
  \\
  &=\frac{1}{N}\int e^{-i(\mathbf{G}+\mathbf{k}_{\mu\nu})\cdot\mathbf{r}}
  \sum_{\mathbf{T}_\mu} e^{-i\mathbf{k}_\mu\cdot\mathbf{T}_\mu} \mu^*(\mathbf{r}-\mathbf{T}_\mu)
  \sum_{\mathbf{T}_\nu} e^{ i\mathbf{k}_\nu\cdot\mathbf{T}_\nu} \nu(\mathbf{r}-\mathbf{T}_\nu) d\mathbf{r}
  \\
  &=\frac{1}{N}\int\sum_{\mathbf{T}_\mu}
  e^{-i(\mathbf{G}+\mathbf{k}_{\mu\nu})\cdot(\mathbf{r}+\mathbf{T}_\mu)}
  e^{-i\mathbf{k}_\mu\cdot\mathbf{T}_\mu} \mu^*(\mathbf{r})
  \sum_{\mathbf{T}_\nu} e^{ i\mathbf{k}_\nu\cdot(\mathbf{T}_\mu+\mathbf{T}_\nu)} \nu(\mathbf{r}-\mathbf{T}_\nu) d\mathbf{r}
  \\
  &=\sum_{\mathbf{T}_\nu} e^{i\mathbf{k}_\nu\cdot\mathbf{T}_\nu}
  \int e^{-i(\mathbf{G}+\mathbf{k}_{\mu\nu})\cdot\mathbf{r}}
  \mu^*(\mathbf{r})\nu(\mathbf{r}-\mathbf{T}_\nu) d\mathbf{r}
\end{align*}

Integral \eqref{eq:vbar} is computed as
\begin{align*}
  \bar{V}_P^\mathbf{k}
  &=\lim_{N\rightarrow\infty}\frac{1}{\sqrt{N}}
  \int \frac{\phi^\mathbf{k}_Q(\mathbf{r}_2)}{r_{12}} d\mathbf{r}_1d\mathbf{r}_2
  \\
  &=\lim_{N\rightarrow\infty}\frac{1}{N} \sum_\mathbf{T}e^{i\mathbf{k}\cdot\mathbf{T}}
  \int \frac{\chi_P(\mathbf{r}_2)-\xi_P(\mathbf{r}_2)}{r_{12}}d\mathbf{r}_1d\mathbf{r}_2
\end{align*}
The limits of this integral are non-vanishing only if $\mathbf{k} = 0$ and the
integrands $\chi_P$ and $\xi_P$ are of $s$-type spherical symmetry
\begin{align*}
  \bar{V}_P^\mathbf{k}
  &=\int\frac{1}{r_{12}}\left[\left( \frac{\alpha_{P_\chi}}{\pi} \right)^{3/2}
  e^{-\alpha_{P_\chi}|\mathbf{r}_2-\mathbf{R}|^2}
  -\left( \frac{\alpha_{P_\xi}}{\pi} \right)^{3/2}
  e^{-\alpha_{P_\xi}|\mathbf{r}_2-\mathbf{R}|^2}\right] d\mathbf{r}_1d\mathbf{r}_2
  \\
  &=\frac{1}{(2\pi)^3}\int\int e^{i\mathbf{G}\cdot\mathbf{r}_1} d\mathbf{r_1}
  \frac{4\pi}{G^2}\left(e^{-\frac{G^2}{4\alpha_{P_\chi}}}e^{-i\mathbf{G}\cdot\mathbf{R}}
  - e^{-\frac{G^2}{4\alpha_{P_\xi}}}e^{-i\mathbf{G}\cdot\mathbf{R}}\right) d\mathbf{G}
  \\
  &=\int\delta(\mathbf{G}) \frac{4\pi}{G^2}
  \left(e^{-\frac{G^2}{4\alpha_{P_\chi}}} - e^{-\frac{G^2}{4\alpha_{P_\xi}}}\right)
  e^{-i\mathbf{G}\cdot\mathbf{R}} d\mathbf{G}
  \\
  &=\lim_{G\rightarrow 0} \frac{4\pi}{G^2} \frac{e^{-\frac{G^2}{4\alpha_{P_\chi}}}
  -e^{-\frac{G^2}{4\alpha_{P_\xi}}}}{e^{i\mathbf{G}\cdot\mathbf{R}}}
  \\
  &=\frac{\pi}{\alpha_{P_\xi}}-\frac{\pi}{\alpha_{P_\chi}}
\end{align*}

\section{Analytical Fourier transformation}
\label{sec:AppendixB}
We applied analytical Fourier transformations in this work to guarantee the
accuracy of the two-electron integrals.  Given Gaussian functions
\begin{align*}
  \mu(\mathbf{r})
  &=C_\mu (x-R_{x\mu})^{m_x} (y-R_{y\mu})^{m_y} (z-R_{z\mu})^{m_z}
  e^{-\alpha_{\mu}|\mathbf{r}-\mathbf{R}_\mu|^2},
  \\
  \nu(\mathbf{r})
  &=C_\nu (x-R_{x\nu})^{n_x} (y-R_{y\nu})^{n_y} (z-R_{z\nu})^{n_z}
  e^{-\alpha_{\nu}|\mathbf{r}-\mathbf{R}_\nu|^2},
\end{align*}
analytical Fourier transformations for the Gaussian function products can be
computed as the products of three Cartesian components
\begin{gather*}
  \int e^{-i\mathbf{G}\cdot\mathbf{r}} \mu^*(\mathbf{r}) \nu(\mathbf{r}) d\mathbf{r}
  = C_\mu C_\nu I_{m_x,n_x}^{x} I_{m_y,n_y}^{y} I_{m_z,n_z}^{z} ,
  \\
  I_{m_x,n_x}^{x}
  =\int e^{-iG_x x} (x-R_{x\mu})^{m_x} e^{-\alpha_{\mu}(x-R_{x\mu})^2}
  (x-R_{x\nu})^{n_x} e^{-\alpha_{\nu}(x-R_{x\mu})^2} dx.
\end{gather*}
Each Cartesian component can be evaluated through the recursive relations
\begin{align*}
  I_{0,0}^{x}
  &=\sqrt{\frac{\pi}{\alpha_\mu+\alpha_\nu}}
  e^{-\frac{\alpha_\mu\alpha_\nu}{\alpha_\mu+\alpha_\nu}(R_{x\mu}-R_{x\nu})^2}
  e^{-\frac{G_x^2}{4(\alpha_\mu+\alpha_\nu)}} e^{-iG_x X_{\mu\nu}}
  \\
  I_{1,0}^{x}
  &=-\left(R_{x\mu}-X_{\mu\nu}+\frac{iG_x}{2(\alpha_\mu+\alpha_\nu)}\right) I_{0,0}^{x}
  \\
  I_{m_x,0}^{x}
  &=\frac{m_x-1}{2(\alpha_\mu+\alpha_\nu)} I_{m_x-2,0}^{x}
  - \left(R_{x\mu}-X_{\mu\nu}+\frac{iG_x}{2(\alpha_\mu+\alpha_\nu)}\right)I_{m_x-1,0}^{x}
  \\
  I_{m_x,n_x}^{x}
  &=I_{m_x+1,n_x-1}^{x} + (R_{x\mu}-R_{x\nu}) I_{m_x,n_x-1}^{x}
\end{align*}
where
\begin{align*}
  X_{\mu\nu}
  &=\frac{\alpha_\mu R_{x\mu}+\alpha_\nu R_{x\nu}}{\alpha_\mu+\alpha_\nu}.
\end{align*}

\end{appendices}

\clearpage
\bibliography{mdf}

\providecommand{\latin}[1]{#1}
\makeatletter
\providecommand{\doi}
  {\begingroup\let\do\@makeother\dospecials
  \catcode`\{=1 \catcode`\}=2\doi@aux}
\providecommand{\doi@aux}[1]{\endgroup\texttt{#1}}
\makeatother
\providecommand*\mcitethebibliography{\thebibliography}
\csname @ifundefined\endcsname{endmcitethebibliography}
  {\let\endmcitethebibliography\endthebibliography}{}
\begin{mcitethebibliography}{36}
\providecommand*\natexlab[1]{#1}
\providecommand*\mciteSetBstSublistMode[1]{}
\providecommand*\mciteSetBstMaxWidthForm[2]{}
\providecommand*\mciteBstWouldAddEndPuncttrue
  {\def\EndOfBibitem{\unskip.}}
\providecommand*\mciteBstWouldAddEndPunctfalse
  {\let\EndOfBibitem\relax}
\providecommand*\mciteSetBstMidEndSepPunct[3]{}
\providecommand*\mciteSetBstSublistLabelBeginEnd[3]{}
\providecommand*\EndOfBibitem{}
\mciteSetBstSublistMode{f}
\mciteSetBstMaxWidthForm{subitem}{(\alph{mcitesubitemcount})}
\mciteSetBstSublistLabelBeginEnd
  {\mcitemaxwidthsubitemform\space}
  {\relax}
  {\relax}

\bibitem[Whitten(1973)]{Whitten1973}
Whitten,~J.~L. \emph{J. Chem. Phys.} \textbf{1973}, \emph{58}, 4496--4501\relax
\mciteBstWouldAddEndPuncttrue
\mciteSetBstMidEndSepPunct{\mcitedefaultmidpunct}
{\mcitedefaultendpunct}{\mcitedefaultseppunct}\relax
\EndOfBibitem
\bibitem[Dunlap \latin{et~al.}(1979)Dunlap, Connolly, and Sabin]{Dunlap1979}
Dunlap,~B.~I.; Connolly,~J. W.~D.; Sabin,~J.~R. \emph{J. Chem. Phys.}
  \textbf{1979}, \emph{71}, 3396--3402\relax
\mciteBstWouldAddEndPuncttrue
\mciteSetBstMidEndSepPunct{\mcitedefaultmidpunct}
{\mcitedefaultendpunct}{\mcitedefaultseppunct}\relax
\EndOfBibitem
\bibitem[Vahtras \latin{et~al.}(1993)Vahtras, Alml\"of, and
  Feyereisen]{Vahtras1993}
Vahtras,~O.; Alml\"of,~J.; Feyereisen,~M. \emph{Chem. Phys. Lett.}
  \textbf{1993}, \emph{213}, 514 -- 518\relax
\mciteBstWouldAddEndPuncttrue
\mciteSetBstMidEndSepPunct{\mcitedefaultmidpunct}
{\mcitedefaultendpunct}{\mcitedefaultseppunct}\relax
\EndOfBibitem
\bibitem[Eichkorn \latin{et~al.}(1995)Eichkorn, Treutler, \"Ohm, H\"aser, and
  Ahlrichs]{Eichkorn1995}
Eichkorn,~K.; Treutler,~O.; \"Ohm,~H.; H\"aser,~M.; Ahlrichs,~R. \emph{Chem.
  Phys. Lett.} \textbf{1995}, \emph{240}, 283 -- 290\relax
\mciteBstWouldAddEndPuncttrue
\mciteSetBstMidEndSepPunct{\mcitedefaultmidpunct}
{\mcitedefaultendpunct}{\mcitedefaultseppunct}\relax
\EndOfBibitem
\bibitem[Dunlap(2000)]{Dunlap2000}
Dunlap,~B.~I. \emph{Phys. Chem. Chem. Phys.} \textbf{2000}, \emph{2},
  2113--2116\relax
\mciteBstWouldAddEndPuncttrue
\mciteSetBstMidEndSepPunct{\mcitedefaultmidpunct}
{\mcitedefaultendpunct}{\mcitedefaultseppunct}\relax
\EndOfBibitem
\bibitem[Beebe and Linderberg(1977)Beebe, and Linderberg]{Beebe1977}
Beebe,~N. H.~F.; Linderberg,~J. \emph{Int. J. Quantum Chem.} \textbf{1977},
  \emph{12}, 683--705\relax
\mciteBstWouldAddEndPuncttrue
\mciteSetBstMidEndSepPunct{\mcitedefaultmidpunct}
{\mcitedefaultendpunct}{\mcitedefaultseppunct}\relax
\EndOfBibitem
\bibitem[R{\o}eggen and Wisl{\o}ff-Nilssen(1986)R{\o}eggen, and
  Wisl{\o}ff-Nilssen]{Roeeggen1986}
R{\o}eggen,~I.; Wisl{\o}ff-Nilssen,~E. \emph{Chem. Phys. Lett.} \textbf{1986},
  \emph{132}, 154 -- 160\relax
\mciteBstWouldAddEndPuncttrue
\mciteSetBstMidEndSepPunct{\mcitedefaultmidpunct}
{\mcitedefaultendpunct}{\mcitedefaultseppunct}\relax
\EndOfBibitem
\bibitem[Aquilante \latin{et~al.}(2009)Aquilante, Gagliardi, Pedersen, and
  Lindh]{Aquilante2009}
Aquilante,~F.; Gagliardi,~L.; Pedersen,~T.~B.; Lindh,~R. \emph{J. Chem. Phys.}
  \textbf{2009}, \emph{130}, 154107\relax
\mciteBstWouldAddEndPuncttrue
\mciteSetBstMidEndSepPunct{\mcitedefaultmidpunct}
{\mcitedefaultendpunct}{\mcitedefaultseppunct}\relax
\EndOfBibitem
\bibitem[F\"usti-Molnar and Pulay(2002)F\"usti-Molnar, and
  Pulay]{Fuesti-Molnar2002}
F\"usti-Molnar,~L.; Pulay,~P. \emph{J. Chem. Phys.} \textbf{2002}, \emph{116},
  7795--7805\relax
\mciteBstWouldAddEndPuncttrue
\mciteSetBstMidEndSepPunct{\mcitedefaultmidpunct}
{\mcitedefaultendpunct}{\mcitedefaultseppunct}\relax
\EndOfBibitem
\bibitem[F\"usti-Molnar(2003)]{Fuesti-Molnar2003}
F\"usti-Molnar,~L. \emph{J. Chem. Phys.} \textbf{2003}, \emph{119},
  11080--11087\relax
\mciteBstWouldAddEndPuncttrue
\mciteSetBstMidEndSepPunct{\mcitedefaultmidpunct}
{\mcitedefaultendpunct}{\mcitedefaultseppunct}\relax
\EndOfBibitem
\bibitem[Lippert \latin{et~al.}(1997)Lippert, Hutter, and
  Parrinello]{Lippert1997}
Lippert,~G.; Hutter,~J.; Parrinello,~M. \emph{Mol. Phys.} \textbf{1997},
  \emph{92}, 477--488\relax
\mciteBstWouldAddEndPuncttrue
\mciteSetBstMidEndSepPunct{\mcitedefaultmidpunct}
{\mcitedefaultendpunct}{\mcitedefaultseppunct}\relax
\EndOfBibitem
\bibitem[Lippert \latin{et~al.}(1999)Lippert, Hutter, and
  Parrinello]{Lippert1999}
Lippert,~G.; Hutter,~J.; Parrinello,~M. \emph{Theor. Chem. Acc.} \textbf{1999},
  \emph{103}, 124--140\relax
\mciteBstWouldAddEndPuncttrue
\mciteSetBstMidEndSepPunct{\mcitedefaultmidpunct}
{\mcitedefaultendpunct}{\mcitedefaultseppunct}\relax
\EndOfBibitem
\bibitem[F\"usti-Molnar and Kong(2005)F\"usti-Molnar, and
  Kong]{Fuesti-Molnar2005}
F\"usti-Molnar,~L.; Kong,~J. \emph{J. Chem. Phys.} \textbf{2005}, \emph{122},
  074108\relax
\mciteBstWouldAddEndPuncttrue
\mciteSetBstMidEndSepPunct{\mcitedefaultmidpunct}
{\mcitedefaultendpunct}{\mcitedefaultseppunct}\relax
\EndOfBibitem
\bibitem[Murphy \latin{et~al.}(2000)Murphy, Cao, Beachy, Ringnalda, and
  Friesner]{Murphy2000}
Murphy,~R.~B.; Cao,~Y.; Beachy,~M.~D.; Ringnalda,~M.~N.; Friesner,~R.~A.
  \emph{J. Chem. Phys.} \textbf{2000}, \emph{112}, 10131--10141\relax
\mciteBstWouldAddEndPuncttrue
\mciteSetBstMidEndSepPunct{\mcitedefaultmidpunct}
{\mcitedefaultendpunct}{\mcitedefaultseppunct}\relax
\EndOfBibitem
\bibitem[Neese \latin{et~al.}(2009)Neese, Wennmohs, Hansen, and
  Becker]{Neese2009}
Neese,~F.; Wennmohs,~F.; Hansen,~A.; Becker,~U. \emph{Chem. Phys.}
  \textbf{2009}, \emph{356}, 98 -- 109\relax
\mciteBstWouldAddEndPuncttrue
\mciteSetBstMidEndSepPunct{\mcitedefaultmidpunct}
{\mcitedefaultendpunct}{\mcitedefaultseppunct}\relax
\EndOfBibitem
\bibitem[Feyereisen \latin{et~al.}(1993)Feyereisen, Fitzgerald, and
  Komornicki]{Feyereisen1993}
Feyereisen,~M.; Fitzgerald,~G.; Komornicki,~A. \emph{Chem. Phys. Lett.}
  \textbf{1993}, \emph{208}, 359 -- 363\relax
\mciteBstWouldAddEndPuncttrue
\mciteSetBstMidEndSepPunct{\mcitedefaultmidpunct}
{\mcitedefaultendpunct}{\mcitedefaultseppunct}\relax
\EndOfBibitem
\bibitem[Komornicki and Fitzgerald(1993)Komornicki, and
  Fitzgerald]{Komornicki1993}
Komornicki,~A.; Fitzgerald,~G. \emph{J. Chem. Phys.} \textbf{1993}, \emph{98},
  1398--1421\relax
\mciteBstWouldAddEndPuncttrue
\mciteSetBstMidEndSepPunct{\mcitedefaultmidpunct}
{\mcitedefaultendpunct}{\mcitedefaultseppunct}\relax
\EndOfBibitem
\bibitem[Weigend \latin{et~al.}(1998)Weigend, Häser, Patzelt, and
  Ahlrichs]{Weigend1998}
Weigend,~F.; Häser,~M.; Patzelt,~H.; Ahlrichs,~R. \emph{Chem. Phys. Lett.}
  \textbf{1998}, \emph{294}, 143 -- 152\relax
\mciteBstWouldAddEndPuncttrue
\mciteSetBstMidEndSepPunct{\mcitedefaultmidpunct}
{\mcitedefaultendpunct}{\mcitedefaultseppunct}\relax
\EndOfBibitem
\bibitem[Sierka \latin{et~al.}(2003)Sierka, Hogekamp, and Ahlrichs]{Sierka2003}
Sierka,~M.; Hogekamp,~A.; Ahlrichs,~R. \emph{J. Chem. Phys.} \textbf{2003},
  \emph{118}, 9136--9148\relax
\mciteBstWouldAddEndPuncttrue
\mciteSetBstMidEndSepPunct{\mcitedefaultmidpunct}
{\mcitedefaultendpunct}{\mcitedefaultseppunct}\relax
\EndOfBibitem
\bibitem[Manby \latin{et~al.}(2001)Manby, Knowles, and Lloyd]{Manby2001}
Manby,~F.~R.; Knowles,~P.~J.; Lloyd,~A.~W. \emph{J. Chem. Phys.} \textbf{2001},
  \emph{115}, 9144--9148\relax
\mciteBstWouldAddEndPuncttrue
\mciteSetBstMidEndSepPunct{\mcitedefaultmidpunct}
{\mcitedefaultendpunct}{\mcitedefaultseppunct}\relax
\EndOfBibitem
\bibitem[Werner \latin{et~al.}(2003)Werner, Manby, and Knowles]{Werner2003}
Werner,~H.-J.; Manby,~F.~R.; Knowles,~P.~J. \emph{J. Chem. Phys.}
  \textbf{2003}, \emph{118}, 8149--8160\relax
\mciteBstWouldAddEndPuncttrue
\mciteSetBstMidEndSepPunct{\mcitedefaultmidpunct}
{\mcitedefaultendpunct}{\mcitedefaultseppunct}\relax
\EndOfBibitem
\bibitem[Sodt \latin{et~al.}(2006)Sodt, Subotnik, and Head-Gordon]{Sodt2006}
Sodt,~A.; Subotnik,~J.~E.; Head-Gordon,~M. \emph{J. Chem. Phys.} \textbf{2006},
  \emph{125}\relax
\mciteBstWouldAddEndPuncttrue
\mciteSetBstMidEndSepPunct{\mcitedefaultmidpunct}
{\mcitedefaultendpunct}{\mcitedefaultseppunct}\relax
\EndOfBibitem
\bibitem[Werner and Manby(2006)Werner, and Manby]{Werner2006}
Werner,~H.-J.; Manby,~F.~R. \emph{J. Chem. Phys.} \textbf{2006}, \emph{124},
  054114\relax
\mciteBstWouldAddEndPuncttrue
\mciteSetBstMidEndSepPunct{\mcitedefaultmidpunct}
{\mcitedefaultendpunct}{\mcitedefaultseppunct}\relax
\EndOfBibitem
\bibitem[Jung \latin{et~al.}(2005)Jung, Sodt, Gill, and Head-Gordon]{Jung2005}
Jung,~Y.; Sodt,~A.; Gill,~P. M.~W.; Head-Gordon,~M. \emph{Proc. Natl. Acad.
  Sci. U. S. A.} \textbf{2005}, \emph{102}, 6692--6697\relax
\mciteBstWouldAddEndPuncttrue
\mciteSetBstMidEndSepPunct{\mcitedefaultmidpunct}
{\mcitedefaultendpunct}{\mcitedefaultseppunct}\relax
\EndOfBibitem
\bibitem[Krack and Parrinello(2000)Krack, and Parrinello]{krack2000all}
Krack,~M.; Parrinello,~M. \emph{Phys. Chem. Chem. Phys.} \textbf{2000},
  \emph{2}, 2105--2112\relax
\mciteBstWouldAddEndPuncttrue
\mciteSetBstMidEndSepPunct{\mcitedefaultmidpunct}
{\mcitedefaultendpunct}{\mcitedefaultseppunct}\relax
\EndOfBibitem
\bibitem[VandeVondele \latin{et~al.}(2005)VandeVondele, Krack, Mohamed,
  Parrinello, Chassaing, and Hutter]{vandevondele2005quickstep}
VandeVondele,~J.; Krack,~M.; Mohamed,~F.; Parrinello,~M.; Chassaing,~T.;
  Hutter,~J. \emph{Comput. Phys. Comm.} \textbf{2005}, \emph{167},
  103--128\relax
\mciteBstWouldAddEndPuncttrue
\mciteSetBstMidEndSepPunct{\mcitedefaultmidpunct}
{\mcitedefaultendpunct}{\mcitedefaultseppunct}\relax
\EndOfBibitem
\bibitem[F\"usti-Molnar and Pulay(2002)F\"usti-Molnar, and
  Pulay]{Fuesti-Molnar2002a}
F\"usti-Molnar,~L.; Pulay,~P. \emph{J. Chem. Phys.} \textbf{2002}, \emph{117},
  7827--7835\relax
\mciteBstWouldAddEndPuncttrue
\mciteSetBstMidEndSepPunct{\mcitedefaultmidpunct}
{\mcitedefaultendpunct}{\mcitedefaultseppunct}\relax
\EndOfBibitem
\bibitem[Dovesi \latin{et~al.}(2005)Dovesi, Civalleri, Roetti, Saunders, and
  Orlando]{Dovesi2005}
Dovesi,~R.; Civalleri,~B.; Roetti,~C.; Saunders,~V.~R.; Orlando,~R. \emph{Rev.
  in Comput. Chem.}; John Wiley \& Sons, Inc., 2005; pp 1--125\relax
\mciteBstWouldAddEndPuncttrue
\mciteSetBstMidEndSepPunct{\mcitedefaultmidpunct}
{\mcitedefaultendpunct}{\mcitedefaultseppunct}\relax
\EndOfBibitem
\bibitem[Kolafa and Perram(1992)Kolafa, and Perram]{Kolafa1992}
Kolafa,~J.; Perram,~J.~W. \emph{Mol. Simul.} \textbf{1992}, \emph{9},
  351--368\relax
\mciteBstWouldAddEndPuncttrue
\mciteSetBstMidEndSepPunct{\mcitedefaultmidpunct}
{\mcitedefaultendpunct}{\mcitedefaultseppunct}\relax
\EndOfBibitem
\bibitem[Gibbon and Sutmann(2002)Gibbon, and Sutmann]{gibbon2002}
Gibbon,~P.; Sutmann,~G. \emph{Lecture Notes, J. Grotendorst, D. Marx, A.
  Muramatsu (Eds.), John von Neumann Institute for Computing, Jülich, NIC
  Series}; 2002; pp 467--506\relax
\mciteBstWouldAddEndPuncttrue
\mciteSetBstMidEndSepPunct{\mcitedefaultmidpunct}
{\mcitedefaultendpunct}{\mcitedefaultseppunct}\relax
\EndOfBibitem
\bibitem[McClain \latin{et~al.}(2017)McClain, Sun, Chan, and
  Berkelbach]{McClain2017}
McClain,~J.; Sun,~Q.; Chan,~G. K.-L.; Berkelbach,~T.~C. \emph{J. Chem. Theory
  Comput.} \textbf{2017}, \emph{13}, 1209--1218\relax
\mciteBstWouldAddEndPuncttrue
\mciteSetBstMidEndSepPunct{\mcitedefaultmidpunct}
{\mcitedefaultendpunct}{\mcitedefaultseppunct}\relax
\EndOfBibitem
\bibitem[Sun \latin{et~al.}(2017)Sun, Berkelbach, Blunt, Booth, Guo, Li, Liu,
  McClain, Sayfutyarova, Sharma, Wouters, and Chan]{pyscf}
Sun,~Q.; Berkelbach,~T.~C.; Blunt,~N.~S.; Booth,~G.~H.; Guo,~S.; Li,~Z.;
  Liu,~J.; McClain,~J.; Sayfutyarova,~E.~R.; Sharma,~S.; Wouters,~S.; Chan,~G.
  K.-L. The Python-based Simulations of Chemistry Framework (PySCF). 2017\relax
\mciteBstWouldAddEndPuncttrue
\mciteSetBstMidEndSepPunct{\mcitedefaultmidpunct}
{\mcitedefaultendpunct}{\mcitedefaultseppunct}\relax
\EndOfBibitem
\bibitem[Goedecker \latin{et~al.}(1996)Goedecker, Teter, and
  Hutter]{Goedecker1996}
Goedecker,~S.; Teter,~M.; Hutter,~J. \emph{Phys. Rev. B} \textbf{1996},
  \emph{54}, 1703\relax
\mciteBstWouldAddEndPuncttrue
\mciteSetBstMidEndSepPunct{\mcitedefaultmidpunct}
{\mcitedefaultendpunct}{\mcitedefaultseppunct}\relax
\EndOfBibitem
\bibitem[Hartwigsen \latin{et~al.}(1998)Hartwigsen, Goedecker, and
  Hutter]{Hartwigsen1998}
Hartwigsen,~C.; Goedecker,~S.; Hutter,~J. \emph{Phys. Rev. B} \textbf{1998},
  \emph{58}, 3641\relax
\mciteBstWouldAddEndPuncttrue
\mciteSetBstMidEndSepPunct{\mcitedefaultmidpunct}
{\mcitedefaultendpunct}{\mcitedefaultseppunct}\relax
\EndOfBibitem
\bibitem[Hutter \latin{et~al.}(2014)Hutter, Iannuzzi, Schiffmann, and
  VandeVondele]{WCMS-CP2K}
Hutter,~J.; Iannuzzi,~M.; Schiffmann,~F.; VandeVondele,~J. \emph{WIREs: Comput.
  Mol. Sci.} \textbf{2014}, \emph{4}, 15--25\relax
\mciteBstWouldAddEndPuncttrue
\mciteSetBstMidEndSepPunct{\mcitedefaultmidpunct}
{\mcitedefaultendpunct}{\mcitedefaultseppunct}\relax
\EndOfBibitem
\end{mcitethebibliography}

\clearpage

\begin{table}
  \centering
  \caption{Even-tempered basis, $\alpha \beta^i, i = 0,\dots,n-1$}
  \begin{tabular}{clllllllll}
    \hline
    angular momentum & $n$ & $\alpha$ & $\beta$ \\
    \hline
    \multicolumn{4}{l}{H atom $10s6p2d$} \\
    $s$ & 10 & 0.244  & 1.6 \\
    $p$ & 6  & 0.596  & 1.6 \\
    $d$ & 2  & 1.454  & 1.6 \\
    \multicolumn{4}{l}{Si atom $20s16p13d7f2g$} \\
    $s$ & 20 & 0.333  & 1.8 \\
    $p$ & 16 & 0.324  & 1.8 \\
    $d$ & 13 & 0.316  & 1.8 \\
    $f$ & 7  & 0.310  & 1.8 \\
    $g$ & 2  & 0.550  & 1.8 \\
    \hline
  \end{tabular}
  \label{tab:auxbasis}
\end{table}


\clearpage

\end{document}